\documentclass[twocolumn,prb,showpacs,preprintnumbers,amsmath,amssymb,superscriptaddress,floatfix]{revtex4}
\usepackage{dcolumn}
\usepackage{bm}
\usepackage{graphicx,epsfig}
\usepackage{bbm}
\usepackage[usenames]{color}
\usepackage{ulem}
\usepackage{slashbox}
\begin{document}

\title{\bf{Real-time density matrix renormalization group dynamics of spin and charge transport 
in push-pull polyenes and related systems}}

\author{Tirthankar Dutta} 
\email{tirthankar@sscu.iisc.ernet.in}
\affiliation{Solid State and Structural Chemistry Unit, Indian Institute of
Science, Bangalore 560012, India}
\affiliation{Condensed Matter Theory Unit, Jawaharlal Nehru Centre for Advanced Scientific Research, Jakkur Post, Bangalore 560064, India.}    
\author{S. Ramasesha}
\email{ramasesh@sscu.iisc.ernet.in}
\affiliation{Solid State and Structural Chemistry Unit, Indian Institute of
Science, Bangalore 560012, India}

\begin{abstract}
In this paper we investigate the effect of terminal substituents on the dynamics of spin and charge 
transport in donor-acceptor substituted polyenes ($D-(CH)_{x}-A$) chains, also known as push-pull polyenes. 
We employ a long-range correlated model Hamiltonian for the $D-(CH)_{x}-A$ system, and  time-dependent density
matrix renormalization group technique for time propagating the wave packet obtained by injecting a hole at 
a terminal site, in the ground state of the system. Our studies reveal that the end groups do not affect spin 
and charge velocities in any significant way, but change the amount of charge transported. We have compared 
these push-pull systems with donor-acceptor substituted polymethine imine (PMI), $D-(CHN)_{x}-A$, systems in 
which besides electron affinities, the nature of $p_{z}$ orbitals in conjugation also alternate from site to 
site. We note that spin and charge dynamics in the PMIs are very different from that observed in the case of 
push-pull polyenes, and within the time scale of our studies, transport of spin and charge leads to the 
formation of a ``quasi-static'' state.
\end{abstract}


\maketitle

\section{INTRODUCTION}
Conjugated organic materials have a variety of applications. They are already being used in organic light emitting 
diodes (OLEDS) and organic thin-film transistors.\cite{torsi,katz,marks,nitzan} They are also considered to be 
potential candidates for single-molecule electronic and spintronic devices. The issue of spin and charge transport in 
$\pi$-conjugated organic systems, therefore, is of prime importance. Thus, it is both important and necessary to 
understand the role of chemical modifications (substituents) on the dynamics of spin and charge transport in these 
systems. Electrons in these molecular materials experience strong electron correlations by virtue of reduced 
dimensionality and these interactions are also long-ranged since the systems are semiconducting. Therefore, to study 
transport of charge and spin in these systems, it is necessary to have appropriate models as well as techniques. 
Dynamics in interacting systems can be studied either in the frequency or in the real-time domain. The dynamics of 
strongly interacting systems in the frequency domain has been possible due to the Lanczos and correction vector 
techniques \cite{karen,ramacv,jeckelmann,whitekuhner} using both exact diagonalization method for small systems, and 
the density matrix renormalization group (DMRG) method for large systems. 

Studying dynamics in the real-time domain throws a lot more light into the transport mechanism. Exact time dependent 
techniques are possible only for small interacting systems. The advent of the 
time-dependent density matrix renormalization 
group (td-DMRG) method has allowed the study of {\it out-of-equilibrium} dynamics in large low-dimensional strongly 
correlated systems.\cite{lxw,whiteTST,schollwockRMP} Recently, we have investigated non-equilibrium dynamics of spin 
and charge transport in unsubstituted polyenes, which are a class of $\pi$-conjugated organic materials, using the 
{\it double time window targeting} (DTWT) td-DMRG technique developed by us.\cite{dutta} In the present paper we extend 
our studies to address non-equilibrium spin and charge dynamics in the {\it push-pull} polyenes, characterized by the 
presence of an electron donating (push) group, and an electron accepting (pull) group, schematically represented as 
$D-(CH)_{x}-A$. Presence of the donor and acceptor groups polarizes the $\pi$-electron bridge such that there is partial 
charge transfer from the donor to the acceptor group. The electronic structure of a push-pull polyene can be described 
by two canonical valence bond (VB) structures, namely, a neutral polyene structure and a zwitterionic structure, also 
known as charge transfer(CT) configuration, where an electron is transferred from {\it D} to {\it A}.\cite{zoppo} This 
leads to the naive expectation that these groups will have significant influence on spin-charge dynamics. The 
{\it push-pull} polyenes have been so far studied mostly in the context of nonlinear optical response.\cite{zoppo,cheng,zyss} 

In this paper we report our time-dependent DMRG studies on the spin and charge transport in push-pull polyenes and 
compare these results with those on polymethine imines which are quasi one-dimensional systems with alternate C and 
N atoms in the conjugation backbone. The organization of the paper is as follows: In the next section we provide 
details about the model Hamiltonian and the computational method used in this study. In Sec. III we present our results 
with discussions. Section IV concludes the paper.

\section{MODEL AND COMPUTATIONAL METHODOLOGY}

The Pariser-Parr-Pople (PPP) Hamiltonian \cite{ppp1} with dimerization and site energies is appropriate for describing 
the low energy physics of $\pi$-conjugated systems. The PPP model Hamiltonian is given by 
\begin{equation}
\begin{split}
\hat{H}_{\text{PPP}} &= \sum_{i=1}^{N-1} \sum_{\sigma} t_{0}[1-(-1)^{i} \delta] (\hat{c}^{\dagger}_{i,\sigma}\hat{c}_{i+1,\sigma} + \text{h.c.})\\
                     &+\sum_{i=1}^{N} \epsilon_{i}\hat{n}_{i} + \sum_{i=1}^{N} \frac{U_{i}}{2}\hat{n}_{i}(\hat{n}_{i}-1) \\ 
                     &+ \sum_{i>j} V_{ij} (\hat{n}_{i}-z_{i})(\hat{n_{j}}-z_{j}).  
\end{split}
\end{equation}
Here, $N$ is the number of carbon atoms in the polyene chain, $\hat{c}^{\dagger}_{i,\sigma}$ ($\hat{c}_{i,\sigma}$) 
creates (annihilates) an electron with spin orientation $\sigma$ in the $p_{z}$ orbital of the $i^{\text{th}}$ carbon 
atom, $t_{0}$ is the average transfer integral and, $\delta$ (0 $\le$ $\delta$ $\le$ 1) is the bond alternation 
parameter. The orbital energy and on-site Coulomb repulsion of the $p_{z}$ orbital on the $i^{\text{th}}$ carbon atom 
are given by $\epsilon_{i}$ and $U_{i}$, respectively and $\hat{n}_{i}$ is the number operator on the $i^{\text{th}}$ 
site. $V_{ij}$ is the inter-site Coulomb repulsion between sites $i$ and $j$, and $z_{i}$ is the on-site chemical 
potential at the $i^{\text{th}}$ site. In case of unsubstituted polyene systems,\cite{rama1} $U_{i}$ = 11.26 eV, 
$\epsilon_{i}$ = 0 and $z_{i}$ = 1, for all sites, $t_{0}$ = $-$2.4 eV and $\delta$ = 0.07. The intersite interaction 
between electrons, $V_{ij}$, is interpolated according to the Ohno scheme,\cite{ohno} between {\it U} for $r$ = 0 and 
$\frac{e^2}{r}$ for $r ~\rightarrow ~\infty$ as,
\begin{equation}
V_{ij} ~=~ 14.397 \biggl[ \biggl(\frac{28.794}{U_{i}+U_{j}} \biggr)^{2} ~+~ r^{2}_{ij} \biggr]^{-1/2}.
\end{equation}
{\noindent
We have used single-bond length of 1.495 \AA, double-bond length of 1.299 \AA, and a bond angle of 120$^{\text{o}}$ 
between 
successive bonds. These parameters have correctly predicted a whole range of properties of the low-lying states of 
conjugated systems in general and polyenes in particular.\cite{rama3,rama4,rama5} When push and pull groups are 
introduced, we assume that only the energies of those $p_{z}$ orbitals to which the push and pull groups are attached, 
change due to inductive effect and all other parameters of the model remain unchanged. The donor group raises the 
orbital energy of the first carbon atom to which it is attached by $+\epsilon_{D}$, while the acceptor group lowers the 
orbital energy of the last carbon atom to which it is attached by $-\epsilon_{A}$, where $\epsilon_{D}$ and 
$\epsilon_{A}$ $>$ 0. We assume that $\epsilon_{D}$ = $-\epsilon_{A}$ (symmetric push-pull polyenes). We have studied 
push-pull polyenes of 30 and 40 carbon atoms with terminal push and pull groups, and have varied the push-pull strength 
$|\epsilon|$. Although, presence of the push and pull groups destroys both the electron-hole and the inversion symmetry,
the total spin invariance of the Hamiltonian, remains preserved.}  

For studying the dynamics of spin and charge transport, an initial wave packet $\mid \psi(0) \rangle$ is 
constructed by annihilating an up spin electron from the first site of a push-pull polyene of $N$ sites, in 
the ground state, $\mid \phi^{0}_{\text{gs}} \rangle$,
\begin{equation}
\mid \psi(0) \rangle = c_{1,\uparrow} \mid \phi_{\text{gs}}^{0} \rangle.
\end{equation}
The wave packet $\mid \psi(0) \rangle$ evolves under the influence of the PPP Hamiltonian [Eq. (1)] following the 
time-dependent Schr"odinger equation and temporal dependence of site charge density $\langle n_{i}(t) \rangle$, and site 
spin density $\langle s_{i}^{z}(t) \rangle$, of this wave packet can be computed as, 
\begin{align}
\langle n_{i}(t) \rangle &= \langle \psi(t) \mid (n_{i,\sigma} + n_{i,-\sigma}) \mid \psi(t) \rangle, \\
\langle s^{z}_{i}(t) \rangle &= \frac{1}{2} \langle \psi(t) \mid (n_{i,\sigma} - n_{i,-\sigma}) \mid \psi(t) \rangle.
\end{align}
Here, $\mid \psi(t) \rangle$ is the wave packet at time $t$. 

Real-time dynamics of the initial wave packet is studied using the DTWT scheme,\cite{dutta} with the following 
specifications: number of density matrix eigenvectors (DMEVs) retained, $m$ = 300, time-step for evolution $\Delta \tau$ 
= 0.0066 fs, total evolution time $T$ = 33.0 fs, the number of time steps in each window is kept at 
130 which corresponds to 0.858 fs, and the number of windows is 39. Although we compute charge (spin) density at all the 
sites, we focus only on the quantities $\langle n_{1}(t) \rangle$ and $\langle n_{L}(t) \rangle$ and, 
$\langle s^{z}_{1}(t) \rangle$ and $\langle s^{z}_{L}(t) \rangle$ at the terminal sites attached to the substituents as 
these are sufficient for our purpose.

Another class of donor-acceptor substituted systems examined are the polymethine imines (PMI), with the molecular 
formula $D-(CHN)_{x}-A$.\cite{PMI1,PMI2,PMI3,PMI4} These systems have alternately donor (C) and acceptor (N) atoms 
in conjugation. The bonding in this polymer corresponds to $\cdots-CH=N-CH=N-\cdots$ and both carbon and nitrogen atoms 
are in $sp^{2}$ hybridization. This system has been studied extensively for linear and non-linear optical properties. 
The nitrogen 2$p_{z}$ orbitals are lower in energy than the carbon 2$p_{z}$ orbitals and intra-orbital repulsion of the 
nitrogen 2$p_{z}$ orbitals are greater than that of the carbon 2$p_{z}$ orbitals, both these can be rationalized on the 
basis of the more compact 2$p$ orbitals in nitrogen compared to carbon. The transfer integrals for $C=N$ and $C-N$ are 
$-$2.767 eV and $-$2.317 eV, respectively; $C=N$ and $C-N$ bond lengths are 1.273 \AA and 1.425 \AA; $U_{C}$ = 11.26 eV, 
$U_{N}$ = 12.34 eV; $\epsilon_{C}$ = 0.0 eV, $\epsilon_{N}$ = $-$2.96 eV. The initial wave packet is constructed as 
before [see Eq. (3)] and time evolved by the DTWT scheme using a smaller time-step of $\Delta \tau$ = 0.0033 fs, 
necessitated by the larger transfer integrals. 

\section{RESULTS AND DISCUSSION}

\begin{figure}[!tbp]
\begin{center}
\epsfig{file=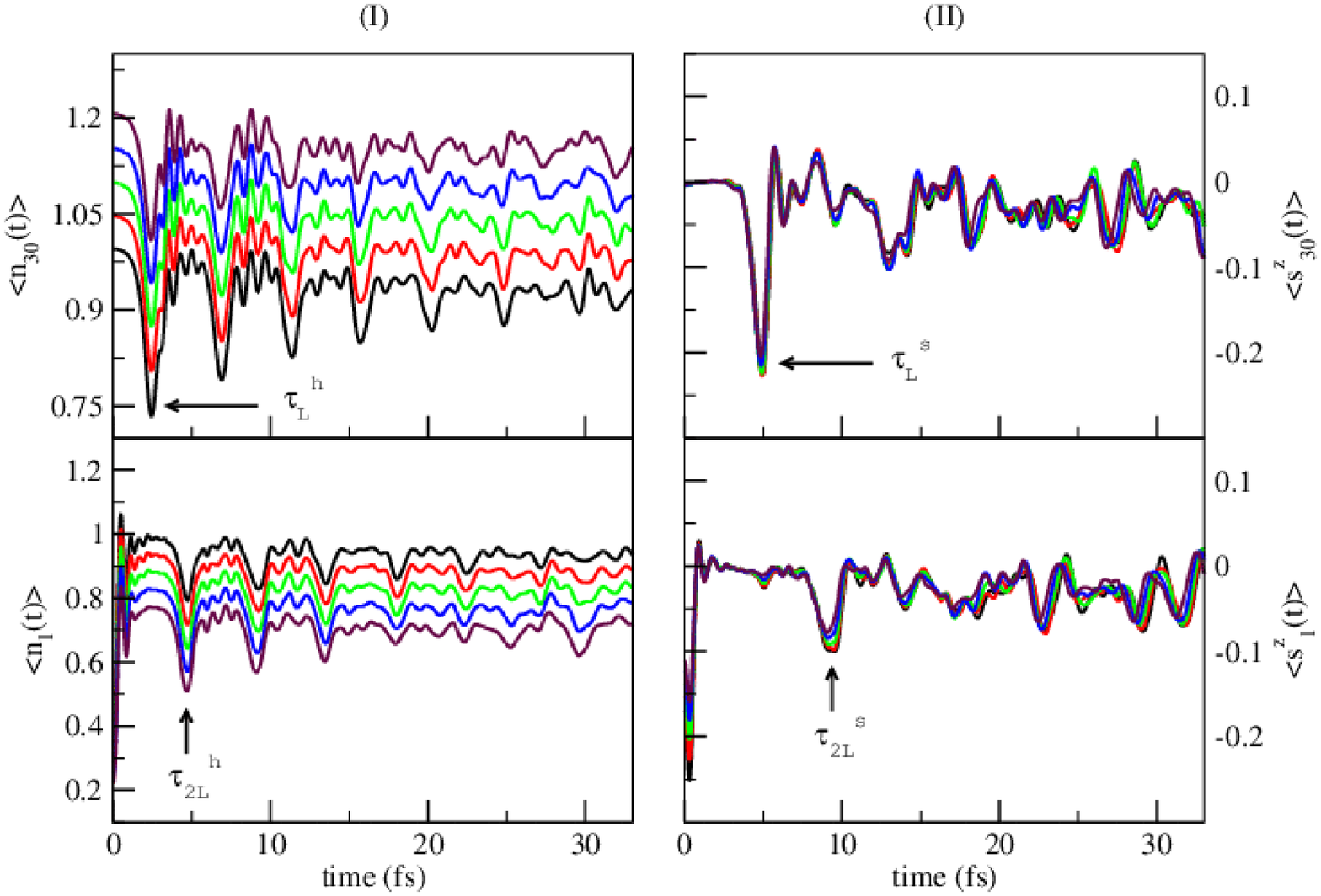,width=3.4 in} \\ 
\vspace*{0.7cm}\epsfig{file=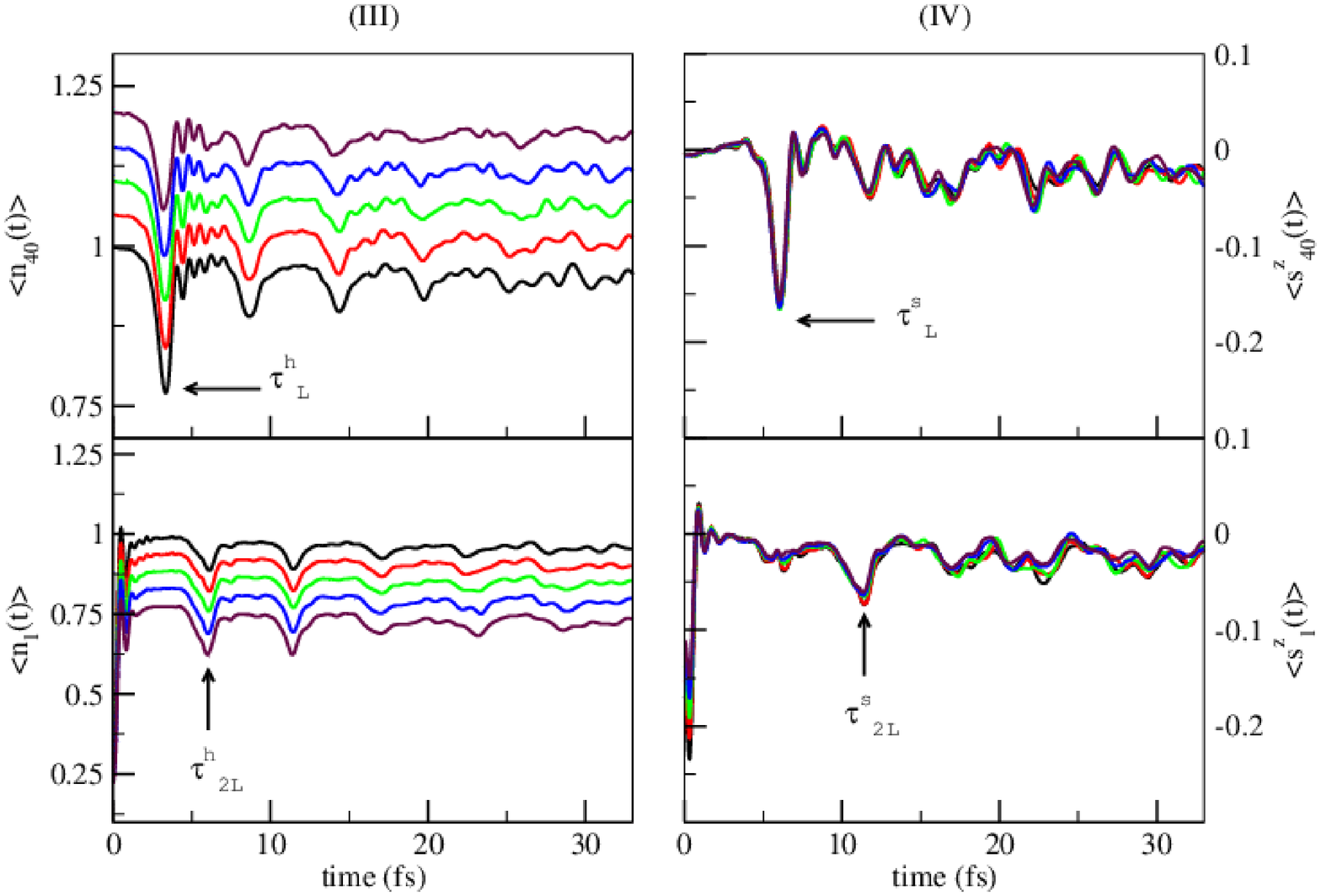,width=3.4 in}
\end{center}
\caption{(color online) \small{Temporal variation of $\langle n_{i}(t) \rangle$ and $\langle s^{z}_{i}(t) \rangle$ 
for dimerized ($\delta$ = 0.07) push-pull polyenes. Panels {\bf (I)} and {\bf (II)} are for 
{\it N} = 30 chains while {\bf (III)} and {\bf (IV)} are for {\it N} = 40 chains. Panels {\bf (I)} and
{\bf (III)} give evolution of charge while {\bf (II)} and {\bf (IV)} give spin density evolution. In each 
panel, bottom box gives the evolution of site {\it 1} and the top box, evolution at the end of the chain
{\it L}. In the charge density panels ({\bf (I)} and {\bf (III)}) $|\epsilon|$ is 0.0 eV, 0.5 eV, 1.0 eV,
1.5 eV and 2.0 eV for curves from top to bottom in the $\langle n_{1}(t) \rangle$ boxes, and from bottom to
top in the $\langle n_{L}(t) \rangle$ boxes. The same is the case in the spin density panels ({\bf (II)} and 
{\bf (IV)}). In color, $|\epsilon|$ = 0.0 eV (black), 0.5 eV (red), 1.0 eV (green), 1.5 eV (blue), 2.0 eV 
(maroon) curves. The positions of $\tau^{h}_{2L}$ and $\tau^{h}_{L}$ in {\bf (I)} and {\bf (III)}, and 
$\tau^{s}_{2L}$ and $\tau^{s}_{L}$ in {\bf (II)} and {\bf (IV)}, are indicated with arrows.}} 
\end{figure}

In the absence of source and sink, the charge and spin of the hole decouple (spin-charge separation) and oscillate 
between sites $1$ and $L$, with time. Hence, the time evolution profiles of $\langle n_{i}(t) \rangle$ and 
$\langle s^{z}_{i}(t) \rangle$ ($i$ = $1$ and $L$), consist of a series of maxima and minima. The time $\tau^{h}_{L}$ 
taken for the charge of the hole, and $\tau^{s}_{L}$ for the spin of the hole, to propagate from site $1$ to site $L$, 
is represented by the time at which the first major minimum (dip) appears in the time evolution profile of 
$\langle n_{L}(t) \rangle$ and $\langle s^{z}_{L}(t) \rangle$, respectively. Time required by the charge degree of 
freedom, $\tau^{h}_{2L}$, and spin degree of freedom, $\tau^{s}_{2L}$, of the hole to travel the round trip from site 
$1$ $\rightarrow$ $L$ $\rightarrow$ $1$, is represented by the time at which the second major minima appears in the time 
evolution profiles of $\langle n_{1}(t) \rangle$ and $\langle s^{z}_{1}(t) \rangle$. From these times, the charge and 
spin velocities of the injected hole are calculated as, $\vartheta^{s/h}_{L}$ = $\frac{L}{\tau^{s/h}_{L}}$ and 
$\vartheta^{s/h}_{2L}$ = $\frac{2L}{\tau^{s/h}_{L}}$.

The dynamics of spin and charge transport in systems with reduced dimensionality and electron-electron interactions, are 
expected to be different from each other, owing to spin-charge separation. When electron correlations are absent, the 
charge and spin velocities, $\vartheta^{h}_{L}$ and $\vartheta^{s}_{L}$, are equal to the Fermi velocity, 
$\vartheta_{F}$. From our earlier td-DMRG studies \cite{duttacomput} we have found that the Fermi velocities of 
tight-binding chains of lengths $30$ and $40$ are 2.63 and 2.72 (\AA/fs) respectively, for $\delta$ = 0.0, and 2.43 and 
2.49 (\AA/fs), respectively, for $\delta$ = 0.07. Table I presents a comparison of the ratio of 
$\vartheta^{h}_{L}/\vartheta_{F}$ and $\vartheta^{s}_{L}/\vartheta_{F}$, for Hubbard chains of 30 and 40 sites with 
different $\frac{U}{|t|}$ values and $\delta$ = 0.07. These quantities are also computed for dimerized PPP chains with 
30 and 40 sites, with $|\epsilon|$ = 0.0, and compared with those of Hubbard chains. As evident from Table I, the values 
of the ratios of $\vartheta^{h}_{L}/\vartheta_{F}$ and $\vartheta^{s}_{L}/\vartheta_{F}$ are much higher in the PPP 
model compared to the Hubbard model, due to the presence of long-range electron-electron interactions. 

\begin{table}[!tbp]
\caption{Variation in $\frac{\vartheta^{h}_{L}}{\vartheta_{F}}$ and $\frac{\vartheta^{s}_{L}}{\vartheta_{F}}$
in the dimerized ($\delta$ = 0.07) Hubbard model with chain lengths 30 and 40, for $\frac{U}{\mid t \mid}$ = 
2.0, 4.0 and 6.0. The same quantities for the $\delta$ = 0.07 PPP model is also quoted. M and Q stand for model and
quantities, respectively.} 
\begin{tabular}{l*{9}{c}}\hline\hline
\backslashbox{Q \kern-0.5em}{M \kern-0.5em} & \multicolumn{2}{c}{$\frac{U}{|t|}$ = $2.0$} & \multicolumn{2}{c}{$\frac{U}{|t|}$ = $4.0$} & \multicolumn{2}{c}{$\frac{U}{|t|}$ = $6.0$} & \multicolumn{2}{c}{PPP} \\ 
\hline
$N$                                   & 30   & 40           & 30   & 40               & 30   & 40               & 30   & 40 \\
&&&&&&&&  \\
$\vartheta^{h}_{L}/\vartheta_{F}$     & 1.33 & 1.32         & 1.35 & 1.39             & 1.30 & 1.18             & 5.05 & 4.80 \\ 
&&&&&&&&  \\
$\vartheta^{s}_{L}/\vartheta_{F}$     & 0.97 & 0.91         & 0.69 & 0.80             & 0.68 & 0.80             & 2.49 & 2.63 \\
&&&&&&&&  \\
\hline\hline
\end{tabular}
\end{table}

As with unsubstituted polyenes,\cite{dutta} in push-pull polyenes (Fig. 1) the spin propagates slower than the charge 
with $\tau^{s}_{L/2L}$ $>$ $\tau^{h}_{L/2L}$ ($\vartheta^{s}_{L/2L}$ $<$ $\vartheta^{h}_{L/2L}$). However, we find that 
the {\it D} and {\it A} groups have no significant effect on spin and charge velocities in the push-pull polyenes. 
Consequently, the charge and spin velocities in push-pull polyenes remain almost equal to those in unsubstituted polyenes 
(Table II). The push-pull substitution however, has an effect on the amount of charge transported, as seen from the 
depths of the minima in the time evolution profiles of $\langle n_{1}(t) \rangle$ and $\langle n_{L}(t) \rangle$. 
However, depth of the minima in $\langle s^{z}_{1}(t) \rangle$ or $\langle s^{z}_{L}(t) \rangle$ do not change (see 
Fig. 1), implying that the push-pull groups have no influence on the spin degree of freedom of the hole.

\begin{table}[!tbp]
\caption{\small{Variation of $\tau^{h}_{L}$ and $\tau^{s}_{L}$ (fs) and, $\vartheta^{h}_{L}$ and $\vartheta^{s}_{L}$ 
(\AA/fs) with chain length ($L$), in symmetric push-pull polyene chains of length, 30 and 40 sites, with different 
values of $|\epsilon|$ (eV). The PPP model parameters are, $t_{0}$ = $-$2.4 eV, $U$ = 11.26 eV, and $\delta$ = 0.07; 
$L$ = 1.397$(N-1-\delta)$ \AA, $N$ being the number of sites, and $O$ stands for Observables.}}
\begin{tabular}{c c c c c c c}\hline\hline
$N$ & $O$ & $|\epsilon|$=0.0 &$|\epsilon|$=0.5 &$|\epsilon|$=1.0 &$|\epsilon|$=1.5 &$|\epsilon|$=2.0 \\
\hline
   & $\tau^{h}_{L}$ & 2.45 & 2.45 & 2.44 & 2.44 & 2.42  \\
   & $\tau^{s}_{L}$ & 4.96 & 4.92 & 4.90 & 4.87 & 4.83 \\
30 & $\vartheta^{h}_{L}$ & 16.50 & 16.50 & 16.57 & 16.57 & 16.71 \\
   & $\vartheta^{s}_{L}$ & 8.15 & 8.22 & 8.25 & 8.30 & 8.37 \\
   & $\biggl(\vartheta^{h}_{L}/\vartheta^{s}_{L}\biggr)$ & 2.02 & 2.01 & 2.01 & 2.00 & 2.00  \\
   & $\tau^{h}_{L}$ & 3.35 & 3.34 & 3.30 & 3.25 & 3.19 \\
   & $\tau^{s}_{L}$ & 6.10 & 6.08 & 6.05 & 6.03 & 6.01 \\
40 & $\vartheta^{h}_{L}$ & 16.24 & 16.28 & 16.42 & 16.74 & 17.05 \\
   & $\vartheta^{s}_{L}$ & 8.92 & 8.95 & 8.99 & 9.02 & 9.05 \\
   &  $\biggl(\vartheta^{h}_{L}/\vartheta^{s}_{L}\biggr)$ & 1.82 & 1.82 & 1.83 & 1.86 & 1.88 \\
\hline\hline
\end{tabular}
\end{table}
This implies that even though the spin and charge velocities in $D-(CH)_{x}-A$ polyenes remain largely unaffected by the 
donor-acceptor strengths, the amount of charge transferred in unit time decreases with increase in $|\epsilon|$ as seen 
from the decrease in depth of the minima, when a hole is doped at the donor site. We can expect the opposite of when we 
dope an electron at the donor site; more charge will be transported from the donor in these cases as $|\epsilon|$ 
increases. Our results indicate that push-pull substitutions do not alter the many-body character of the $D-(CH)_{x}-A$ 
polyenes, and that spin and charge velocities are controlled by the correlation strength while the magnitude of charge 
transfer is controlled by the strength of the substituents.

To understand in detail, the reason for the inability of the donor and acceptor groups to alter the 
many-body character of the $D-(CH)_{x}-A$ polyenes, we compute the inverse of the charge gap ($\Delta E_{c}$) 
and spin gap ($\Delta E_s$) of the push-pull polyenes. If $E_{gs}(N+1)$, $E_{gs}(N-1)$ and $E_{gs}(N)$ 
denote the ground state energies of the $(N+1)$-particle, $(N-1)$-particle and $N$-particle systems 
respectively, then the charge gap is given by,
\begin{equation}
\Delta E_{c} = E_{gs}(N+1) + E_{gs}(N-1) - 2 E_{gs}(N).
\end{equation}
Similarly, the spin gap is defined as,
\begin{equation}
\Delta E_{s} = E_{0}(S^{z}_{\text{tot}}=3/2) - E_{0}(S^{z}_{\text{tot}}=1/2),
\end{equation}
where, $E_{0}(S^{z}_{\text{tot}}=3/2)$ and $E_{0}(S^{z}_{\text{tot}}=1/2)$ are the lowest energy states of the 
$D-(CH)_{x}-A$ 
polyenes with $S^{z}_{\text{tot}}$ $3/2$ and $1/2$, respectively. In Fig. 2, we have plotted the variation in 
$\tau_{L}^{h/s}$ and $1/\Delta E_{c/s}$ as a function of the strength of the push-pull groups. It is clearly observed 
that the donor-acceptor groups fail to alter either of the gaps, as a result of which, the velocities of the spin and 
charge remain unaffected.

\begin{figure}[!tbp]
\begin{center}
\epsfig{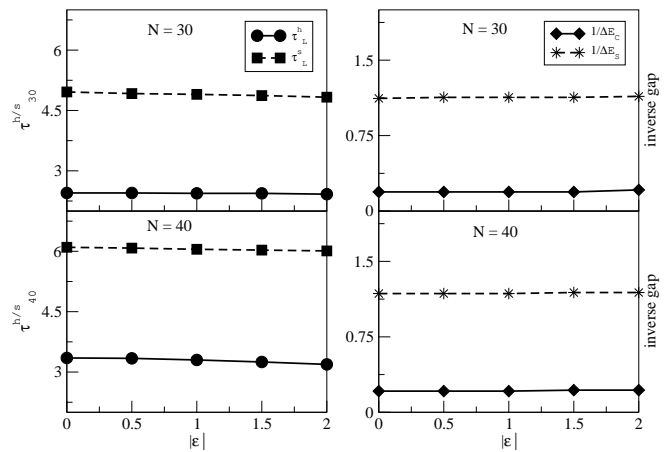}  
\end{center}
\caption{\small{Variation in $\tau^{h/s}_{L}$ (fs), left-hand top and bottom plots, and $1/\Delta E_{c/s}$ (eV$^{-1}$), 
right-hand top and bottom plots, as a function of strength of push-pull groups ($|\epsilon|$ = 0.0, 0.5, 1.0, 1.5 and 
2.0), for polyene chains of 30 and 40 sites.}}
\end{figure}

\begin{figure}[!tbp]
\begin{center}
\epsfig{file=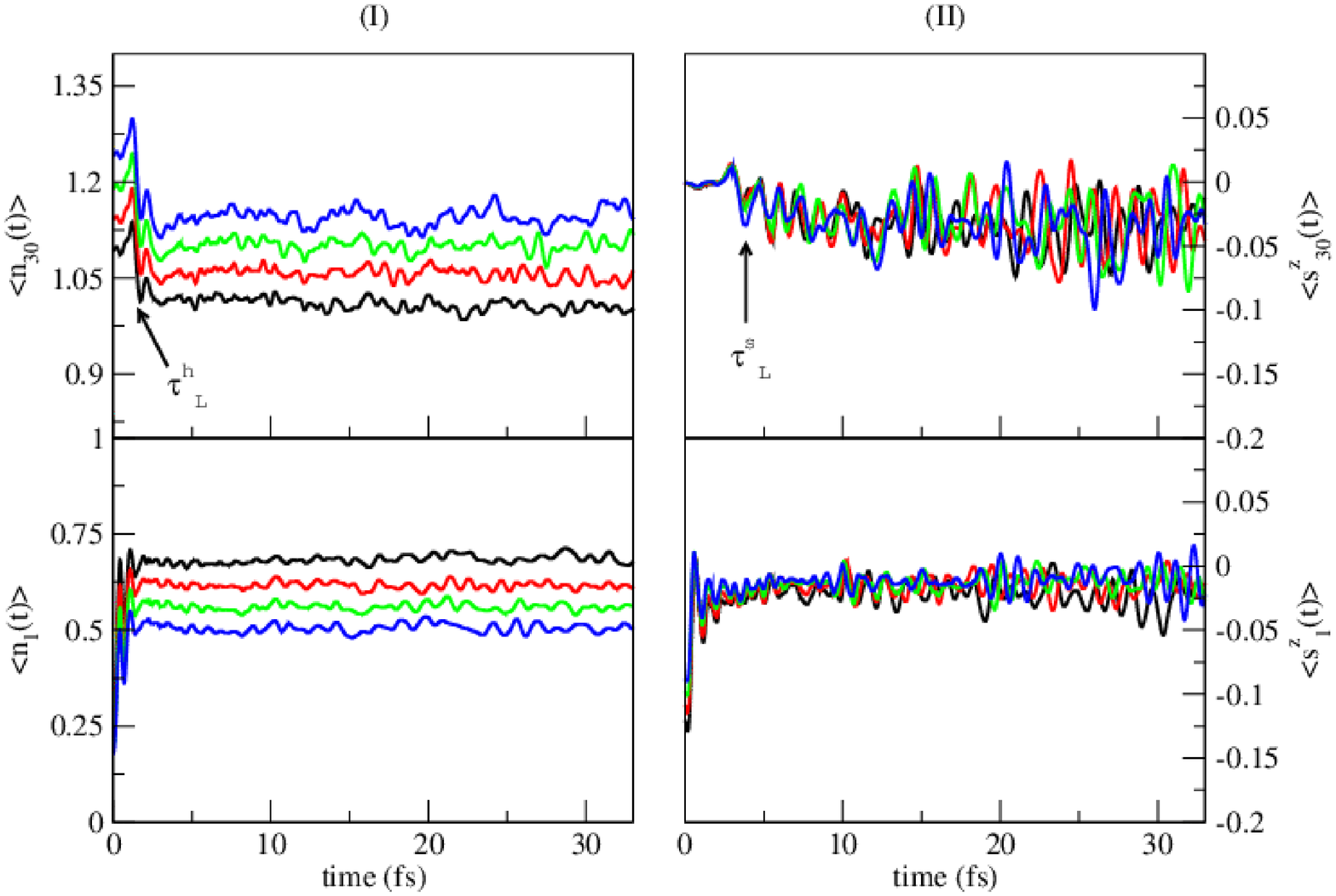,width=3.4 in} \\
\vspace*{0.7cm}\epsfig{file=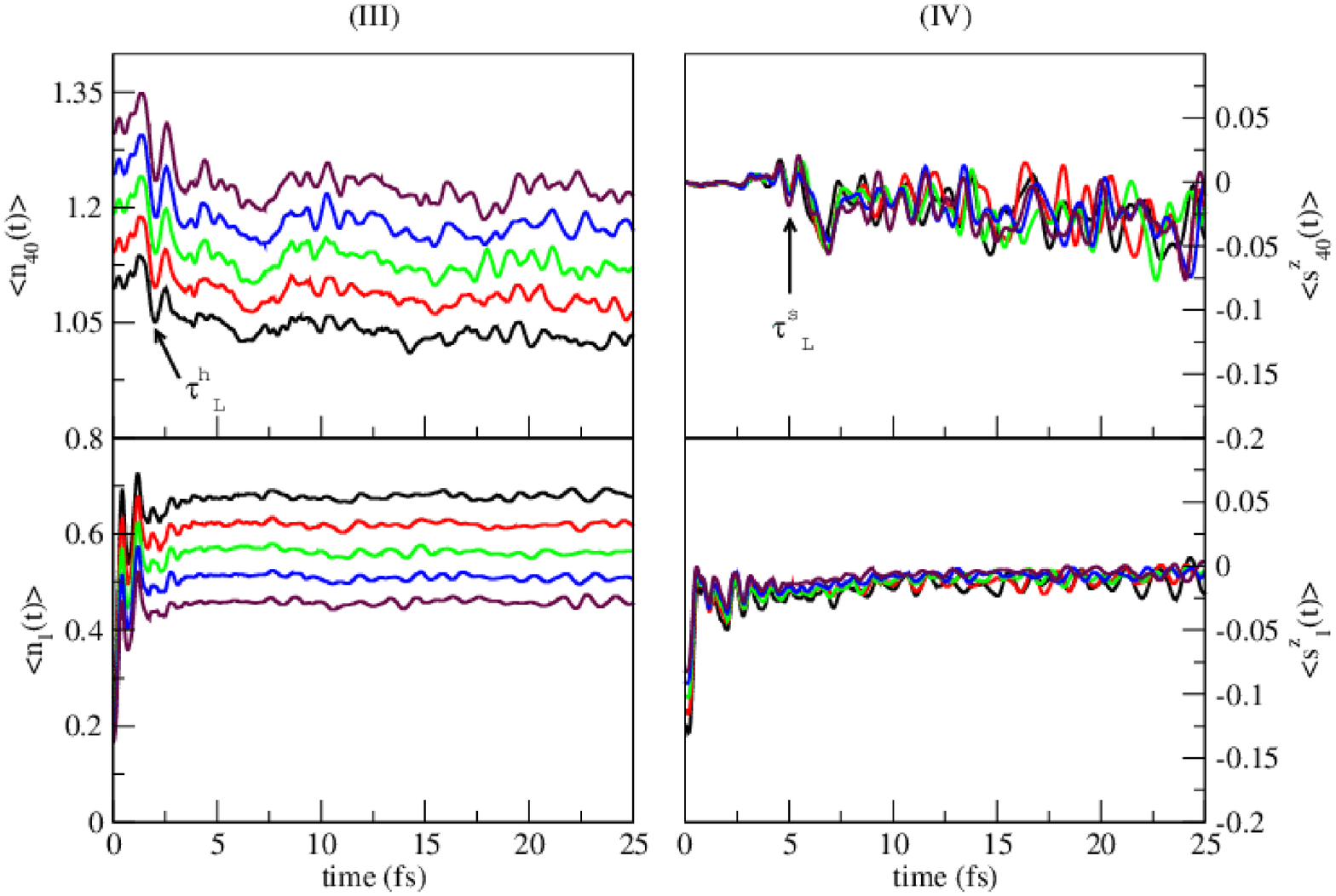,width=3.4in}
\end{center}
\caption{(color online) \small{
Temporal variation in $\langle n_{i}(t) \rangle$ and $\langle s^{z}_{i}(t) \rangle$ for polymethine imine 
system. Panels {\bf (I)} and {\bf (II)} are for $N$ = 30 chains while {\bf (III)} and {\bf (IV)} are for $N$ 
= 40 chains. Panels {\bf (I)} and {\bf (III)} give evolution of charge while {\bf (II)} and {\bf (IV)} give 
spin density evolution. In each panel, bottom box gives the evolution of site $1$ and the top box, evolution 
at the end of the chain $L$. In the charge density panels ({\bf (I)} and {\bf (III)}) $|\epsilon|$ is 0.0 eV,
0.5 eV, 1.0 eV, 1.5 eV and 2.0 eV for curves from top to bottom in the $\langle n_{1}(t) \rangle$ boxes, and 
from bottom to top in the $\langle n_{L}(t) \rangle$ boxes. The same is the case in the spin density panels 
({\bf (II)} and {\bf (IV)}). In color, $|\epsilon|$ = 0.0 eV (black), 0.5 eV (red), 1.0 eV (green), 1.5 eV 
(blue), 2.0 eV (maroon) curves. In top boxes of the panels {\bf (I)}, {\bf (II)}, {\bf (III)} and {\bf (IV)},
$\tau^{h}_{L}$ and $\tau^{s}_{L}$ have been indicated by black arrow.}}
\end{figure}

Comparing Figs. 1 and 3 it is clear that the dynamics of spin and charge transport in the $D-(CHN)_{x}-A$ chains is very 
different from that in the push-pull polyenes, owing to the different $U$ values and site energies of carbon and 
nitrogen atoms. For example, unlike in push-pull polyenes (where the charge and spin of the injected hole after 
decoupling, oscillate back and forth between sites $1$ and $L$) in the $D-(CHN)_{x}-A$ systems, the oscillations are 
damped and the spin and charge of the injected hole do not return after reaching the chain end. This is demonstrated by 
the observation that, although the first dip in the time evolution profiles of $\langle n_{L}(t) \rangle$ and 
$\langle s^{z}_{L}(t) \rangle$ are well defined, there are no well defined second (major) minima in the 
$\langle n_{1}(t) \rangle$ and $\langle s^{z}_{1}(t) \rangle$ curves as a function of time. This may be because the 
hopping of an electron between unlike atoms is not a degenerate resonant tunneling process and hence, is not strictly 
reversible. 

In order to investigate the fate of the injected hole, we study the temporal variation of total charge and spin 
densities in the left and right halves of PMI chains of 30 and 40 sites, for $|\epsilon|$ = 0.0 (Fig. 4). The total 
charge and spin densities for the left (right) half are defined as, $\langle N_{L(R)}(t) \rangle$ = 
$\sum_{j \in L(R)}$ $\langle n_{j}(t) \rangle$, and $\langle S^{z}_{L(R)}(t) \rangle$ = 
$\sum_{j \in L(R)}$ $\langle s^{z}_{j}(t) \rangle$ ($L$ $\in$ $[1, N/2]$, $R$ $\in$ $[N/2+1, N]$), respectively. These 
quantities serve as simple but effective probes to understand the motion of charge and spin of the hole, as they travel 
from the half of the system where charge injection occured to the opposite half of the system. If the time evolution 
profiles of these observables show significant oscillatory behavior, it implies that the charge and spin propagate back 
and forth between the ends of the system. However, if the oscillatory behavior is not pronounced and both the observables 
attain some average value, it signifies that the charge (spin) moves in a such a manner that a ``quasi-static'' state is 
generated in which charge (spin) distribution of the system remains unchanged within the time of study. It is observed 
from Fig. 4 that for both the chain lengths, $\langle S^{z}_{R}(t) \rangle$ goes from 0.0 to $-$0.25 while, 
$\langle S^{z}_{L}(t) \rangle$ goes from an initial value of $-$0.5 to $-$0.25. Once these quantities have attained the 
value of $-$0.25, they start to oscillate with time about this average value. The charge densities for the left and right 
halves of the PMI chains are however, found to behave little differently. In the case of the 30 site chain it is 
observed that, $\langle N_{L}(t) \rangle$ and $\langle N_{R}(t) \rangle$ take the average values of $\sim$14.25 and 
$\sim$14.75, while for the longer chain of 40 sites, these quantities oscillate about a mean value of $\sim$19.5. These 
data suggest that in the $(CHN)_{x}$ systems, with increase in system size, within the computational time a 
``quasi-static'' state is achieved more rapidly.

\begin{figure}[!tbp]
\begin{center}
\epsfig{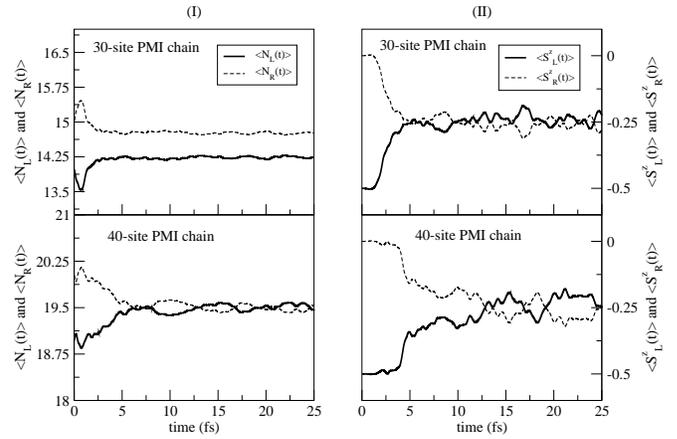} 
\end{center}
\caption{\small{Variation in $\langle N_{L(R)}(t) \rangle$ (plots in left box marked as {\bf (I)}) and 
$\langle S^{z}_{L(R)}(t) \rangle$ (plots in right box marked as {\bf (II)}) with time, for unsubstituted PMI chains 
($|\epsilon|$ = 0.0) of 30 and 40 sites. Solid curves correspond to $\langle N_{L}(t) \rangle$ and 
$\langle S^{z}_{L}(t) \rangle$, respectively, and dashed curves correspond to $\langle N_{R}(t) \rangle$ and 
$\langle S^{z}_{R}(t) \rangle$. }}
\end{figure}

To understand this ``quasi-static'' state in more detail, we plot the time evolution profiles of charge and spin 
densities on carbon and nitrogen atoms in both halves of unsubstituted PMI chains of 30 and 40 sites (Fig. 5). These 
quantities are defined as, 
\begin{eqnarray}
\langle N_{L,C}(t)\rangle = \sum_{j \in C} \langle n_{j}(t) \rangle;~~\langle S^{z}_{L,C}(t)\rangle = \sum_{j \in C} \langle s^{z}_{j}(t) \rangle, \\ 
\langle N_{R,C}(t)\rangle = \sum_{j \in C} \langle n_{j}(t) \rangle;~~\langle S^{z}_{R,C}(t)\rangle = \sum_{j \in C} \langle s^{z}_{j}(t) \rangle, \\ 
\langle N_{L,N}(t)\rangle = \sum_{j \in N} \langle n_{j}(t) \rangle;~~\langle S^{z}_{L,N}(t)\rangle = \sum_{j \in N} \langle s^{z}_{j}(t) \rangle, \\ 
\langle N_{R,N}(t)\rangle = \sum_{j \in N} \langle n_{j}(t) \rangle;~~\langle S^{z}_{R,N}(t)\rangle = \sum_{j \in N} \langle s^{z}_{j}(t) \rangle, \\ 
\end{eqnarray}
where $L$ $\in$ $[1, N/2]$ and $R$ $\in$ $[N/2+1, N]$. It is observed from Figs. 5(II) and 5(IV) that the oscillations 
in time, of $\langle S^{z}_{L/R,C}(t)\rangle$ and $\langle S^{z}_{L/R,N}(t)\rangle$, are opposite in phase, signifying 
that spin densities on the carbon and nitrogen atoms (in both half) are antiferromagnetically coupled to each other. 
However, the total spin density carried by carbon atoms in both halves is found to be less than that carried by the 
nitrogen atoms. Furthermore it is seen that with time, both $\langle S^{z}_{L,C}(t)\rangle$ and 
$\langle S^{z}_{L,N}(t)\rangle$ decrease, while $\langle S^{z}_{R,C}(t)\rangle$ and $\langle S^{z}_{R,N}(t)\rangle$ 
increase in magnitude, keeping the total spin density in either half at $-$0.25. From Figs. 5(I) and 5(III) we see that 
the average value of $\langle N_{L/R,C}(t)\rangle$ is less than $\langle N_{L/R,N}(t)\rangle$, and the total charge 
density carried by nitrogen atoms on the right half of PMI chains is more compared to that in the left half in line 
with the fact that the nitrogens are the ``acceptor'' atoms, being more electronegative. The opposite is seen for the 
carbon atoms. The time evolution profiles of total spin densities on carbon and nitrogen in both halves, compared to the 
total charge densities, are found to be more oscillatory.

\begin{figure}[!tbp]
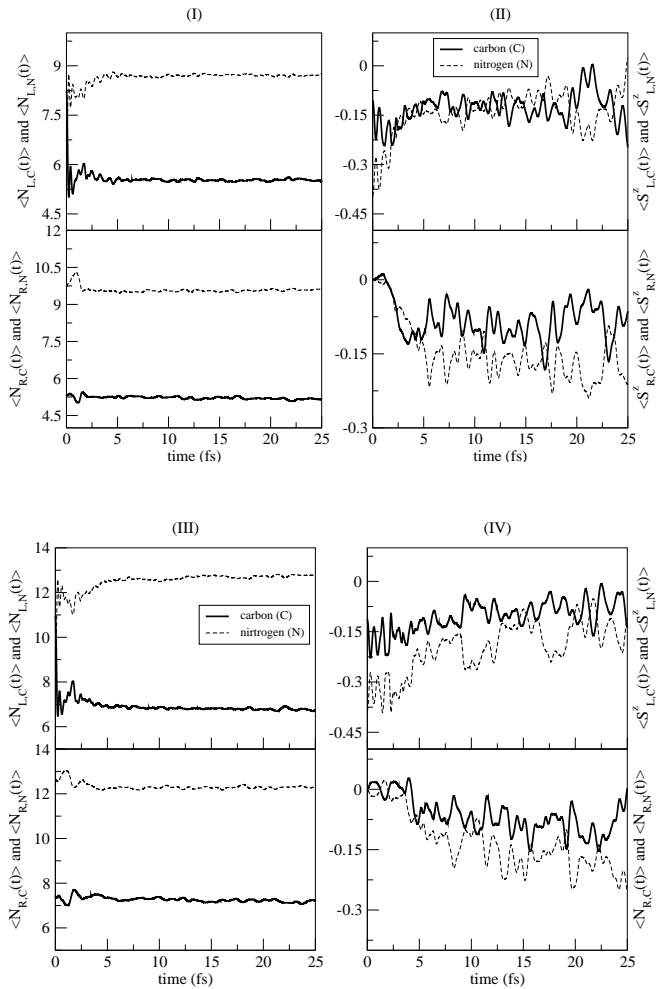

\begin{center}
\epsfig{file=Tirthankar_fig7.eps,width=3.4 in} \\
\vspace*{0.7cm}\epsfig{file=Tirthankar_fig8.eps,width=3.4in}
\end{center}
\caption{\small{Variation in total charge and spin densities of carbon and nitrogen atoms, with time, in the left and 
right halves of unsubstituted PMI chains ($|\epsilon|$ = 0.0) with 30 and 40 sites. Top and bottom boxes correspond to 
$N$ = 30 and 40 sites, respectively. Solid and dashed curves pertain to carbon and nitrogen, respectively. In the top 
box, plots marked {\bf (I)} refer to charge density variation and plots marked {\bf (II)} refer to spin density 
variation, in left (upper plots) and right (lower plots) half of the system, respectively. The same holds for plots 
marked as {\bf (III)} and {\bf (IV)}, in the bottom box.}}
\end{figure}

All these observations suggest that the ``quasi-static'' state is characterized by charge and spin of the hole 
distributed between the left and right halves of the chain in almost equal amount, the positive or hole charge 
predominantly residing on carbon atoms while the spin, on nitrogen atoms. The $N$ atoms being more electronegative than 
the $C$ atoms ($\epsilon_{N}$ = $-$2.96 eV, $\epsilon_{C}$ = 0.0 eV), prefer being electron rich by accommodating more 
charge density. However, due to large on-site Coulomb repulsion ($U_{N}$ = 12.34 eV), the average site charge on $N$ 
atoms is slightly larger than 1.0 which is observed from the charge density distribution in the neutral ground state and 
the initial state. On the other hand, due to $U_{C}$ = 11.26 eV and $\epsilon_{C}$ = 0.0, the carbon atoms are slightly
electron deficient in the $(CHN)_{x}$ system. 
Thus, propagation of the charge and spin degrees of freedom of the hole in this 
polarized background results in formation of the ``quasi-static'' state, which once formed, prevents both the charge 
and spin from returning to the site of injection as it requires reversal of polarization and is hence, energetically 
unfavorable. 

In unsubstituted PMI chains,  $\tau^{h}_{L}$ and $\tau^{s}_{L}$ are 1.75 fs and 3.65 fs for $N$ = 30, and 2.01 fs and 
6.60 fs for $N$ = 40. We find that the $\vartheta_{h}/\vartheta_{s}$ ratios for the $N$ = 30 is 2.08 and for $N$ = 40, 
3.29. The $\vartheta_{h}/\vartheta_{s}$ ratio for a PMI chain of 30 sites agrees well with that of push-pull polyene 
with $N$ = 30 ($|\epsilon|$ = 0). However, this ratio for $N$ = 40 ($|\epsilon|$ = 0) is much higher in the PMI chain 
compared to push-pull polyene of same size. This is opposite to the behavior exhibited by push-pull polyenes, where the 
$\vartheta_{h}/\vartheta_{s}$ ratio decreases with increasing chain length. In both polyenes and polymethine imines, the
 $\vartheta_{h}/\vartheta_{s}$ ratios do not saturate, indicating that the $\pi$-coherence lengths for transport are 
rather long. It also appears that alternate donor-acceptor sites along the chain enhance the velocities of both charge 
and spin. We have also studied substituents at the terminal sites in the $(CHN)_{x}$ systems and as with the polyenes, 
the strength of the push-pull groups do not affect the charge and spin velocities, or the nature of the ``quasi-static'' 
state.

\section{CONCLUSIONS}

To conclude, our studies show that the velocities of charge and spin transport are not affected by push-pull 
substituents, although the magnitude of charge transport depends on the push-pull strength. In the related $(CHN)_{x}$ 
system we find that, due to the underlying polarized structure both the spin and charge transport are affected, and 
charge travels much faster than spin compared to push-pull systems. Furthermore, the transport of charge and spin 
results in the formation of a ``quasi-static'' state in which the injected hole resides in both halves of the system, 
almost in equal amount. We also note that in the PMIs the finite size effects are very large compared to the push-pull
polyenes, suggesting much longer $\pi$-coherence lengths.

\begin{center}
{\bf ACKNOWLEDGMENTS}
\end{center}
This work was supported by DST India and the Swedish Research Link Program under the Swedish Research Council.

\end{document}